\documentclass[twocolumn,english,aps,prl,groupedaddress,superscriptaddress]{revtex4}
\usepackage{graphicx,epsfig,units}
\usepackage{xcolor} 
\usepackage{soul} 
\usepackage{amsmath,amsfonts,mathrsfs,amsbsy,bm,babel}

\begin{document}

\title{Two-dimensional magnetic interactions in LaFeAsO}


\author{M.~Ramazanoglu}
\affiliation{Ames Laboratory, Ames, IA, 50011, USA}
\affiliation{Department of Physics and Astronomy, Iowa State University, Ames, IA, 50011, USA}

\author{J.~Lamsal}
\affiliation{Ames Laboratory, Ames, IA, 50011, USA}
\affiliation{Department of Physics and Astronomy, Iowa State University, Ames, IA, 50011, USA}

\author{G.~S.~Tucker}
\affiliation{Ames Laboratory, Ames, IA, 50011, USA}
\affiliation{Department of Physics and Astronomy, Iowa State University, Ames, IA, 50011, USA}

\author{J.-Q.~Yan}
\affiliation{Oak Ridge National Laboratory, Oak Ridge, TN, 37831, USA}

\author{S.~Calder}
\affiliation{Oak Ridge National Laboratory, Oak Ridge, TN, 37831, USA}

\author{T.~Guidi}
\affiliation{ISIS Facility, Rutherford Appleton Laboratory, Chilton, Didcot, Oxon OX11 OQX, United Kingdom}

\author{T.~Perring}
\affiliation{ISIS Facility, Rutherford Appleton Laboratory, Chilton, Didcot, Oxon OX11 OQX, United Kingdom}

\author{R.~W.~McCallum}
\affiliation{Ames Laboratory, Ames, IA, 50011, USA}
\affiliation{Department of Physics and Astronomy, Iowa State University, Ames, IA, 50011, USA}

\author{T.~A.~Lograsso}
\affiliation{Ames Laboratory, Ames, IA, 50011, USA}
\affiliation{Department of Physics and Astronomy, Iowa State University, Ames, IA, 50011, USA}

\author{A.~Kreyssig}
\affiliation{Ames Laboratory, Ames, IA, 50011, USA}
\affiliation{Department of Physics and Astronomy, Iowa State University, Ames, IA, 50011, USA}

\author{A.~I.~Goldman}
\affiliation{Ames Laboratory, Ames, IA, 50011, USA}
\affiliation{Department of Physics and Astronomy, Iowa State University, Ames, IA, 50011, USA}

\author{R.~J.~McQueeney}
\affiliation{Ames Laboratory, Ames, IA, 50011, USA}
\affiliation{Department of Physics and Astronomy, Iowa State University, Ames, IA, 50011, USA}

\begin{abstract}
Inelastic neutron scattering measurements demonstrate that the magnetic interactions in antiferromagnetic LaFeAsO are two-dimensional. 
Spin wave velocities within the Fe layer and the magnitude of the spin gap are similar to the \textit{A}Fe$_2$As$_2$ based materials. 
However, the ratio of interlayer and intralayer exchange is found to be less than $\sim 10^{-4}$ in LaFeAsO, very similar to the cuprates, and $\sim$ 
100 times smaller than that found in \textit{A}Fe$_2$As$_2$ compounds. The results suggest that the effective dimensionality of the magnetic system is 
highly variable in the parent compounds of the iron arsenides and weak 3-D interactions may limit the maximum attainable superconducting $T_{c}$.
\end{abstract}

\date{March 15,2013}
\maketitle

The discovery of high-temperature superconductivity in the iron arsenide compounds \cite{Kamihara08} immediately led to comparisons to the copper oxide 
superconductors. Both systems possess layered crystal structures, suggesting that two-dimensional (2-D) behavior may be a shared feature amongst the 
high-temperature superconductors.  In particular, the enhanced spin fluctuations that arise from reduced dimensionality is regarded as a critical element 
of high-temperature superconductivity.  In the case of the copper oxide materials, the magnetism occurs within square copper oxide sheets that are weakly 
coupled to each other due to separation by ionic layers (such as BaO or LaO).  The parent La$_{2}$CuO$_{4}$ compound, for example, is an insulator with a 
strongly anisotropic resistivity measured within ($\rho_{ab}$) and perpendicular ($\rho_{c}$) to the Cu layers ($\rho_{c}/\rho_{ab} \approx $ 500 at high 
temperatures).\cite{Preyer89}  The magnetic excitations measured with inelastic neutron scattering (INS) 
are well-understood within a 2-D Heisenberg model as the ratio of interlayer to intralayer exchange is very small ($J_{c}/J_{ab} 
\approx 10^{-4}-10^{-5}$).\cite{Endoh88,Keimer92}

In the iron arsenides, magnetism also occurs in separated square FeAs layers, but the dimensionality of the magnetic interactions is debated. 
Measurements of the anisotropic properties have been mainly performed on the \textit{A}Fe$_{2}$As$_{2}$ (122) system ($A=$ Ca, Sr, Ba), where large 
single-crystals are available.  In the 122 materials, the FeAs layers are separated by an alkali-earth metal layer (with Fe-Fe layer separation of 5.5 - 6.5 \AA\ 
from Ca - Ba, respectively). Transport properties in parent 122  compounds display only a weak anisotropy (for example, $\rho_{c}/\rho_{ab} \approx 1-3$ 
\cite{Tanatar09}) in contradiction to the strongly 2-D transport properties observed in the cuprates.  Angle-resolved photoemission 
(ARPES) experiments on the 122 compounds indicate a significant variation of the Fermi surface geometry along the $c$-axis that is also consistent
with a 3-D system.\cite{Heimes, Kim}  Finally, INS measurements with $A=$ Ca,\cite{Diallo08, Diallo09, Zhao09}, Sr \cite{Zhao08}, and Ba 
\cite{Sato09, Harriger11}, indicate a fairly substantial interlayer magnetic exchange interaction ($J_{c}/J_{ab} \approx 2-6\%$) which supports 
three-dimensional (3-D) magnetism.

Very little is known about the magnetic interactions in the $R$FeAsO (1111) family of superconducting materials that currently claim the largest superconducting 
transition temperature of $T_{c}^{max}\approx$ 55 K (whereas $T_{c}^{max}\approx$ 40 K for the 122 compounds).\cite{Johnston10} 
Based on a larger interlayer spacing ($\approx$ 8.7 \AA), 1111 compounds are expected to be closer to the 2-D limit than the 122 compounds.  
The recent availability of large single-crystals of LaFeAsO \cite{Yan09} have enabled measurements of the anisotropic resistivity 
($\rho_{c}/\rho_{ab} \approx$ 2-20)\cite{Jesche12}, which is similar to the 122 compounds.  However, ARPES measurements of the 
effective dimensionality of the electronic system are inconclusive due to the presence of surface states.\cite{Liu10}  
In this Letter, we use INS measurements of the spin-wave spectrum in the parent LaFeAsO compound to show that magnetic exchange coupling is 2-D,
despite the inference of only weak anisotropy from bulk measurements, with a ratio of exchange interactions comparable to the cuprates ($J_{c}/J_{ab} < 10^{-4}$).  
This result provides evidence that the magnetism can vary from 2-D to weakly 3-D in different iron arsenide compounds, with possible implications for the 
maximum achievable $T_{c}$.\cite{Monthoux}

The sample used for INS experiments consists of dozens of small single-crystals of LaFeAsO with a total mass of approximately 600 mg that are co-aligned 
to within $\sim$2 degrees.  Details of crystal growth and characterization are described elsewhere.\cite{Yan09} Previous neutron and x-ray scattering 
measurements show that the crystals undergo a tetragonal-orthorhombic structural phase transition at $T_{\mbox{\scriptsize S}}=155$\,K, followed by 
stripe antiferromagnetic ordering transition at $T_{\mbox{\scriptsize N}}=140$\,K.\cite{Yan09} 
The wavevector of the stripe AFM ordered state is $\mathbf{Q}_{AFM}=$ (1/2,1/2,1/2)$_{T}$ when indexed with reference to the high-temperature $P4/nmm$ tetragonal
structure. In this paper, the scattering data is presented with respect to low-tempereture orthorhombic $Cmma$ unit cell (in other words; $\mathbf{Q}_{AFM}=$ (1,0,1/2)$_{O}$)
where we define $\mathbf{Q} = (H,K,L) = \frac{2\pi}{a}H\hat{\imath} + \frac{2\pi}{b}K\hat{\jmath} + \frac{2\pi}{c}L\hat{k}$ and the lattice constants are 
$a \approx b =$ 5.68 \AA\ 
and $c=8.75$ \AA. INS measurements were performed on the MERLIN spectrometer at the ISIS Neutron Scattering Facility at Rutherford-Appleton Laboratory and 
the HB3 spectrometer at the High Flux Isotope Reactor at Oak Ridge National Laboratory.  
For these measurements, the samples were mounted in the $(H,0,L)$ scattering plane.

For subsequent discussion, both the MERLIN and HB3 data are described using a model of damped Heisenberg spin waves with nearest ($J_{1a}$,$J_{1b}$) and 
next-nearest ($J_{2}$) interactions within the Fe layer, and an interlayer exchange ($J_{c}$).  We also include a single-ion anisotropy energy ($D$) to 
account for an observed spin gap.  Within linear spin-wave theory, the dispersion is given by $\hbar\omega(\textbf{q})=\sqrt{A_\textbf{q}^2-B_\textbf{q}^2}$ 
with 
$A_{\textbf{q}}=2S[D+2J_{2}+J_{1a}+J_{c}+J_{1b}(\cos(\pi K)-1)]$ and $B_{\textbf{q}}=2S[J_{1a}\cos(\pi H)+2J_{2}\cos(\pi H)\cos(\pi K)+J_{c}\cos(2\pi L)]$.  
The magnetic susceptibility can be written as a damped simple harmonic oscillator (with damping parameter $\Gamma$),

\begin{equation}
\chi^{''}(\textbf{q},E)=\frac{\chi_{0}\Gamma E}{[E^2-(\hbar\omega(\textbf{q}))^2]^{2}+\Gamma^2E^2}.
\label{eqn1}
\end{equation}
and the INS intensity in arbitrary units is $f^{2}(Q)\chi^{''}(\textbf{Q}-\textbf{Q}_{AFM},E)(1-e^{-E/kT})^{-1}$ where $f(Q)$ is the magnetic form factor 
of the Fe$^{2+}$ ion and $\textbf{q}=\textbf{Q}-\textbf{Q}_{AFM}$ is the reduced wavevector within the magnetic Brillouin zone.

\begin{figure}
\includegraphics[width=0.90\linewidth]{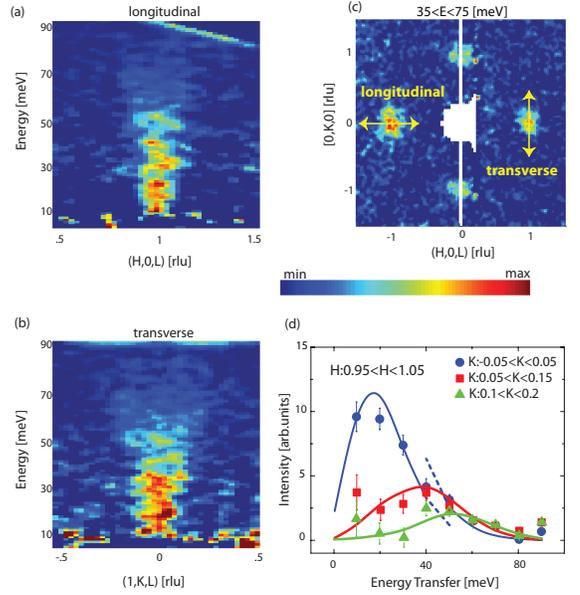}
\caption{\footnotesize (a) Longitudinal and (b) transverse cuts through the inelastic neutron scattering spectrum of LaFeAsO at T=5 K as measured on the 
MERLIN spectrometer with $E_{i}=$150 meV after background substruction (see text).  (c) Data averaged over an energy transfer range from 35-75 meV 
showing anisotropic spin fluctuations in 
the $H-K$ plane centered at $\textbf{Q}_{AFM}$. (d) Energy spectra at different average $K$ values along the transverse direction; 0, 0.1, and 0.15.  
Lines shown are fits described in the text. In these panels, the $L$ component of the wavevector varies with the in-plane wavevector and energy trasfer 
because of the fixed crystal orientation with respect to the incident beam direction.}
\label{fig1}
\end{figure}

MERLIN measurements were performed with the incident neutron beam oriented along $L$ and an incident energy $E_{i}=$ 150 meV. 
The spectrum of spin fluctuations were measured deep in the stripe AFM ordered state at $T=$ 5 K and are shown in Figs. \ref{fig1} (a) and (b) for the 
longitudinal ($H$) and transverse ($K$) directions relative to $\textbf{Q}_{AFM}$ (see Fig. 1(c) for reference).  In these spectra, an assumed isotropic 
and energy-dependent non-magnetic background signal was estimated by summing data at all scattering angles after masking the INS signal near the magnetic 
zone centers.  The spectrum below 100 meV consists of steep spin waves concentrated close to $\textbf{Q}_{AFM}$ and all symmetrically equivalent wavevectors 
in the twinned orthorhombic structure [see Fig. 1(c)].  The signal above $\approx$ 100 meV becomes too weak to observe.  The MERLIN data is therefore best 
understood in the small-$q$ limit with spin waves described by an anisotropic linear dispersion relation,

\begin{equation}
\hbar\omega(\textbf{q})=\sqrt{\Delta_{0}^{2}+v_{a}^{2}q_{x}^{2} + v_{b}^{2}q_{y}^{2}+v_{c}^{2}q_{z}^{2}}.  
\label{eqn2}
\end{equation}
The spin gap and spin wave velocities are given by $\Delta_{0}=2S\sqrt{2DJ_{+}}$, $v_{a}=aSJ_{+}$, $v_{b}=bS\sqrt{J_{+}J_{-}}$, $v_{c}=cS\sqrt{J_{c}J_{+}}$, 
respectively, where $J_{+}=2J_{2}+J_{1a}+J_{c}$ and $J_{-}=2J_{2}-J_{1b}$.  

The anisotropy of the dispersion \textit{within the Fe layer} appears as the elliptical shape of the neutron intensity in Fig. 1(c).  
Estimates of the longitudinal and transverse spin wave velocities (based on fits discussed below) are $v_{a}=$ 555 $\pm$ 100 meV \AA\ and $v_{b}=$ 420 
$\pm$ 55meV \AA\, respectively. 
The in-plane spin wave velocities in LaFeAsO are comparable, though slightly larger, than the 122 materials, in agreement with first-principles 
electronic band structure calculations.\cite{Han09}  The longitudinal velocity exceeds the transverse velocity ($v_{a} > v_{b}$) and the anisotropy of the 
spin excitations within the Fe layer is defined as $\eta = (v_{a}^{2}-v_{b}^{2})/(v_{a}^{2}+v_{b}^{2})$.\cite{Tucker12}  The value of $\eta=$ 0.25 for 
LaFeAsO is similar to the values found in the parent 122 compounds where $\eta=$ 0.2-0.4.\cite{Tucker12}

MERLIN measurements cannot ascertain any substantial interlayer exchange interactions, which should appear as $L$-dependent oscillations in the intensity.  
In time-of-flight INS experiments with the $c$-axis fixed along the incident beam, $L(E)$ is a function of the energy transfer $E$. So, the absence of 
substantial $E$-dependent intensity oscillations in Figs. 1(a), (c), and (d) indicates 2-D magnetism.  

In order to verify this 2-D behavior, we performed measurements of the low-energy spin excitations at $T=$ 5 K on the same sample using the HB3 spectrometer. 
Figure 2(a) shows that a substantial spin gap is observed at $\textbf{Q}_{AFM}=(1,0,1/2)$.  The gap shows an onset of $\approx$ 5 meV and a peak 
at $\approx$ 11 meV and is comparable to the spin gaps observed in $A$Fe$_2$As$_2$ \cite{Diallo08,Zhao08,Sato09,Park12} and NaFeAs \cite{Park12} compounds.  
Low-energy INS measurements on polycrystalline LaFeAsO observe a similar sized spin gap and report a 2D-like response.\cite{Ishikado09}

The difference between the spin gaps at the magnetic zone center $\textbf{Q}_{AFM}$ ($\Delta_{0}$) and the magnetic zone boundary point at 
$\textbf{Q}_{ZB}=(1,0,1)$ ($\textbf{q}_{ZB}=(0,0,1/2)$) provides a direct measurement of $J_{c}$.  Using the Heisenberg model, the difference in spin gaps 
is

\begin{equation}
[\hbar\omega(\textbf{q}_{ZB})]^{2} - \Delta_{0}^{2}=16S^{2}J_{c}J_{+}.
\label{eqn3}
\end{equation}

\begin{figure}
\includegraphics[width=0.9\linewidth]{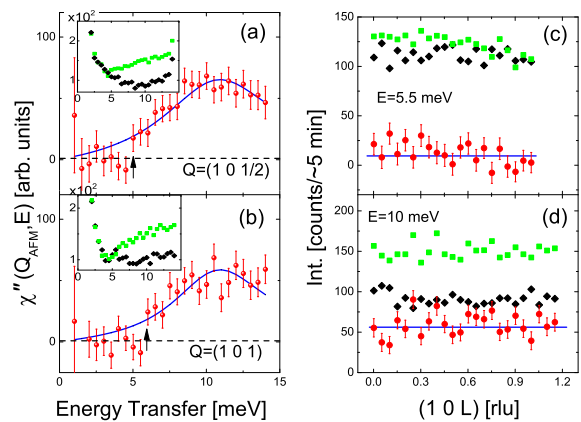}
\caption{\footnotesize  Measurements of the low energy spin excitations in LaFeAsO at $T=$5 K as measured on HB3.   
Energy dependence of the magnetic scattering at (a) $\textbf{Q}_{AFM} =$ (1,0,1/2) and (b) magnetic zone boundary position $\textbf{Q}_{ZB}=(1,0,1)$. 
Constant energy scans depicting the $L$ dependence of the magnetic scattering along (1,0,$L$) (c) at gap onset at 5.5 meV, and (d) above the gap at 10 meV.
Lines are fits to a damped spin wave model as described in the text.
Insets and panels (c) and (d) show the raw data (green squares) and background scans (crystal rotated from nominal $\textbf{Q}$ by 20 degrees, black diamonds) 
that were used to estimate the magnetic scattering (red symbols) in (a)-(d). The arrows in these panels are indicating the onset value of the enegrgy gap.}
\label{fig2}
\end{figure}

\begin{figure}
\includegraphics[width=0.9\linewidth]{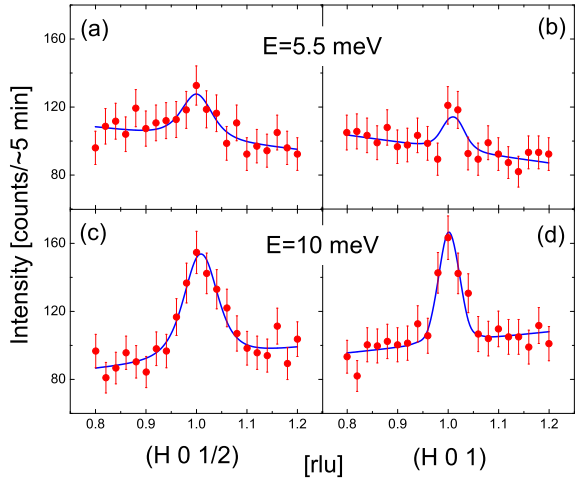}
\caption{\footnotesize  Longitudinal cuts of the low energy spin excitations in LaFeAsO at $T=$ 5 K as measured on HB3. 
Data are shown (a) at the onset of the gap at $E=$ 5.5 meV and $\textbf{Q}_{AFM}=(1,0,1/2)$, (b) at 5.5 meV and $\textbf{Q}_{ZB}=(1,0,1)$, (c) at 10 meV and 
$\textbf{Q}_{AFM}$, and (d) at 10 meV and $\textbf{Q}_{ZB}$. Lines show the fits to the damped spin wave model,described in the text, with only a free amplitude 
and background. The rest of the parameters are fixed to the values given in Table \ref{tab:tablo1}. The quality of the agreement between data and the model
is also a confirmation of the results.}
\label{fig3}
\end{figure}

A comparison of Figs. 2(a) and 2(b) shows that the magnitude of the spin gap is similar at $\textbf{Q}_{ZB}$, thereby providing very strong confirmation that
$J_{c}$ must be weak.  Figs. 2(c) and (d) show an absence of any $L-$dependent sinusoidal modulations of the intensity along $(1,0,L)$ both at the gap onset 
at 5.5 meV and the peak at 10 meV.  Finally, we show in Fig.\ref{fig3} that a longitudinal cut at the gap onset of 5.5 meV reveals weak intensity at both $\textbf{Q}_{AFM}$ 
and $\textbf{Q}_{ZB}$, which again confirms that the two spin gaps are the same. The HB3 data shown in Figs.\ref{fig2} and \ref{fig3} was fit using 
Eqns. \eqref{eqn1} and \eqref{eqn2} after convolution with the instrumental resolution using the RESLIB program\cite{Reslib}.  
These data were used to obtain values for the low energy damping and the spin gaps at $\textbf{Q}_{AFM}$ and $\textbf{Q}_{ZB}$, yielding $\Gamma=8 \pm 1$ meV,
$\Delta_{0}=11.6 \pm 0.5$ meV, and $\hbar\omega(\textbf{q}_{ZB})=11.2 \pm 0.5$ meV, respectively.  All of these data and subsequent fits give 
substantive proof that no observable dispersion exists along $L$ and, therefore, magnetic interactions in LaFeAsO are 2-D in nature.

\begin{figure}
\includegraphics[width=0.9\linewidth]{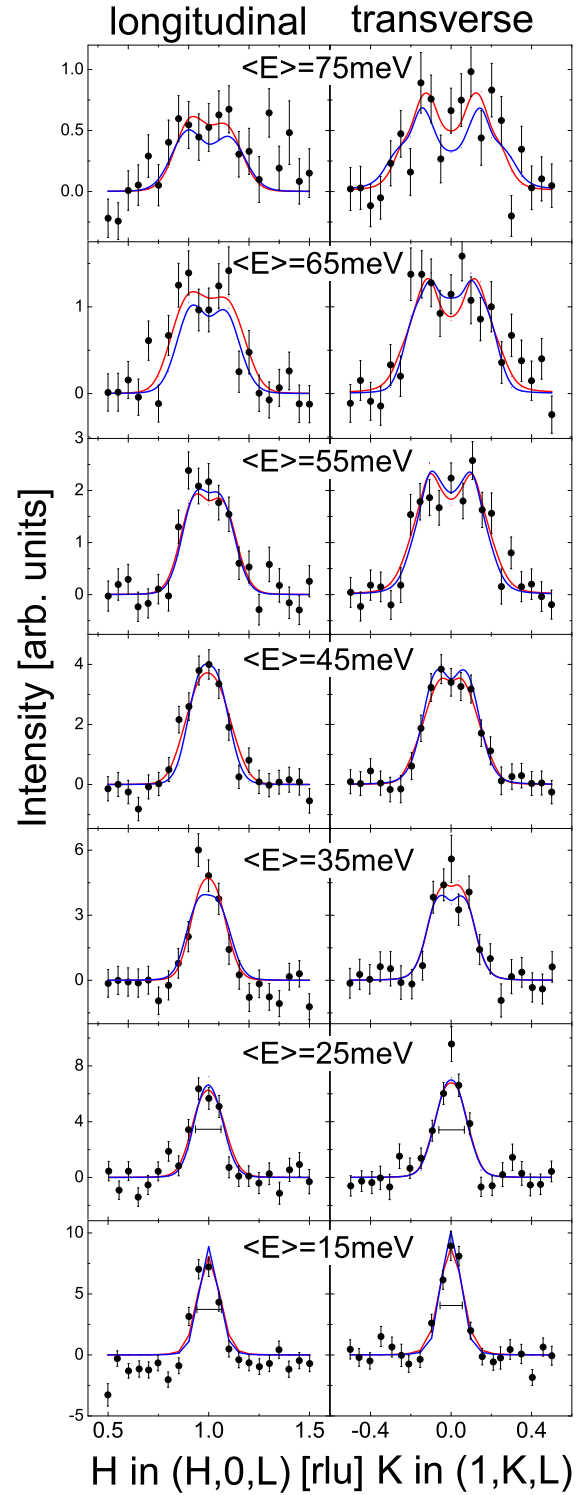}
\caption{\footnotesize  Longitudinal (left) and transverse (right) constant energy cuts of the magnetic scattering in LaFeAsO up to 75 meV as measured on 
MERLIN.   Global (blue line) and local (red line) fits to the damped spin wave model described in the text are shown. The shoulders in the tails
of transverse $E=$75 meV cut fit is due to the effect of the twinning of orthorhombic cyrstal, which is included in the model calculations. For these cuts
the L component of the wavevecotor is function a of the in-plane momentum vectors and the energy transfer, as explained earlier.}
\label{fig4}
\end{figure}

For the MERLIN data, the TOBYFIT program \cite{Tobyfit} was used to fit the corrected data to the 2-D Heisenberg model after convolution with the instrumental 
resolution and accounting for orthorhombic twinning.  The values of $\Delta_{0}$ and $\Gamma$ used in the MERLIN fits are fixed to the values determined 
by the HB3 data. The corrected data has been symmetrized by averaging all four equivalent quadrants of reciprocal space and subtracting an estimate of the 
non-magnetic and background scattering.  The main MERLIN fitting results are displayed in Figs. \ref{fig1} and \ref{fig4} as a series of longitudinal and transverse cuts 
through $\textbf{Q}_{AFM}$ at different energy transfers from $E=$15 to 75 meV.  At low energies, cuts through the steep magnetic excitations consist of a 
single sharp peak centered at $\textbf{Q}_{AFM}$ due to resolution limitations (the resolution width is indicated by the horizontal line in Fig. \ref{fig4}).
Above $E\sim$55 meV, the peak splitting from counter propagating spin wave modes can be resolved and is more pronounced in transverse cuts where the spin 
wave velocity is lower.  The two curves in Fig. \ref{fig4} represent a global fit to all cuts shown (blue line) as well as local fits to each cut (red line) 
with both procedures yielding similar values for the fitting parameters. 
Without the ability to observe the spin wave dispersion at the magnetic zone boundary 
positions, such as $\textbf{q}=(0,1,0)$ [$\textbf{Q}=(1,1,0)$], the fits are not sensitive to the difference between nearest-neighbor exchange constants, $S(J_{1a}-J_{1b})$.  
The full set of fitting parameters listed Table \ref{tab:tablo1} represent both the HB3 and MERLIN data quite well.

\begin{table}
\caption {Parameters obtained from fitting the $J_{1a}-J_{1b}-J_{2}$ spin wave model with damping and single-ion anisotropy.  
The damping factor ($\Gamma$), energy gap ($\Delta_{0}$), and exchange energies $<SJ_{\pm}>$ are in meV while spin-wave velocities 
$v_{a \above 0pt b}$are in meV \AA. }
\label{tab:tablo1}  
\begin{tabular}{|l|l|l|}
~~~~HB3~~~~                     &~~~~ MERLIN Local~~~~     &~~~~MERLIN Global~~~~\\ \hline
$\Gamma=8\pm$1       &$<SJ_{+}>=102\pm$20&$<SJ_{+}>=93\pm$15  \\ 
$\Delta_{0}=11.2\pm$.6  &$<SJ_{-}>=59\pm$7  &$<SJ_{-}>=51\pm$5   \\ \hline
\end{tabular}
\begin{tabular}{|l|}                       
~~~~~~~~~~~$<v_{a}>=555\pm$100~~~~~~~~~~~\\  
~~~~~~~~~~~$<v_{b}>=420\pm$55~~~~~~~~~~~\\              
\hline
\end{tabular}\\
\end{table}

In summary, the details of the spin wave spectrum and magnetic energy scales of LaFeAsO are similar in many ways to the 122 compounds.  
The energy scale for exchange interactions within the Fe layer and their average in-plane anisotropy are nearly equivalent. This is in accordance with 
\textit{ab initio} calculations of the spin excitation spectrum.\cite{Han09} The measurements must be extended up to higher energies in order to 
determine whether any substantial difference exists between $J_{1a}$ and $J_{1b}$.  At lower energies, we find that the magnitude of the spin gap is 
also similar to the 122 compounds.  The common energy scale of the 122 and NaFeAs spin gaps was recently discussed in \cite{Park12}, as it does not follow 
from the expectations of simple single-ion anisotropy due to substantial differences in the magnitude of the ordered moments in the two systems.  
The similar spin gap observed in LaFeAsO, along with its relatively small ordered moment (0.4 $\mu_{B}$),\cite{Cruz08} would seem to add some strength to 
this argument.  

The most important difference in the 122 and 1111 compounds is the interlayer exchange. In our measurements, the zone center and (0,0,1/2) zone 
boundary spin gaps in LaFeAsO are equal within error. Considering the error bars may allow a 1 meV difference in spin gaps, we can estimate an 
upper limit for the exchange anisotropy (based on Eq. \ref{eqn3}) of $J_{c}/J_{+}<10^{-4}$ which is similar to the cuprates and places LaFeAsO strongly 
in the 2-D limit. In comparison, $J_{c}/J_{+} =  2-6\%$ is $\approx$ 100 times larger for the parent 122 
compounds.\cite{Diallo08, Diallo09, Zhao09,Zhao08,Sato09, Harriger11} 
The 2-D antiferromagnetism found in the 1111 compounds may be responsible for some enhancement of ($T_{c}^{max} \approx$ 55 K) and weak 3-D magnetic 
interactions present in the 122 family compounds may present a limitation to higher superconducting transition temperatures ($T_{c}^{max} \approx$ 40 K).

RJM would like to thank D. C. Johnston and V. Antropov for useful discussions. The work at Ames Laboratory was supported by the U.S. Department of Energy, 
Office of Basic Energy Science, Division of Materials Sciences and Engineering under Contract No. DE-AC02-07CH11358. Work at Oak Ridge National Laboratory 
is supported by U.S. Department of Energy, Office of Basic Energy Sciences, Scientific User Facilities Division.

\end{document}